\documentclass[submission]{eptcs}

\usepackage{graphicx}
\usepackage{pdflscape}
\usepackage{afterpage}
\usepackage{caption}
\usepackage{tabularx}
\usepackage{footnote}
\usepackage{url}
\usepackage{amsmath}
\usepackage[T1]{fontenc}
\usepackage{color}
\usepackage{booktabs}
\usepackage{wrapfig}
\usepackage{cite}
\usepackage{hyperref}

\title{Analysis of Feature Models Using Alloy: A Survey}
\makesavenoteenv{table}

\author{Anjali Sree-Kumar, Elena Planas and Robert Claris\'{o}
\email{\{asree\_kumar,eplanash,rclariso\}@uoc.edu} \institute{Universitat Oberta de Catalunya, Barcelona (Spain)}
}

\def\titlerunning{
Analysis of Feature Models Using Alloy: A Survey}
\def\authorrunning{A. Sree-Kumar, E. Planas \& R. Claris\'{o}}
\begin{document}

\maketitle

\begin{abstract}

Feature Models (FMs) are a mechanism to model variability among a family of closely related software products, i.e. a software product line (SPL). Analysis of FMs using formal methods can reveal defects in the specification such as inconsistencies that cause the product line to have no valid products. 

A popular framework used in research for FM analysis is Alloy, a light-weight formal modeling notation equipped with an efficient model finder. Several works in the literature have proposed different strategies to encode and analyze FMs using Alloy. However,  there is little discussion on the relative merits of each proposal, making it difficult to select the most suitable encoding for a specific analysis need. In this paper, we describe and compare those strategies according to various criteria such as the expressivity of the FM notation or the efficiency of the analysis. This survey is the first comparative study of research targeted towards using Alloy for FM analysis. 

This review aims to identify all the best practices on the use of Alloy, as a part of a framework for the automated extraction and analysis of rich FMs from natural language requirement specifications. 

\textbf{Keywords.} Feature model analysis; Alloy; Software Product Line; Formal verification
\end{abstract}

\section{Introduction}  \label{sec:Intro}

Many software systems are not one-of-a-kind, but a family of related products in a given application domain called \emph{software product line} (SPL).  Products in a SPL can be differentiated by their \emph{features}, defined as ``increments in program functionality'' \cite{Batory2006} or ``user-visible aspects or characteristics of the domain'' \cite{Kang1990}. Hence, modeling a SPL requires describing the features and the \emph{relationships} among them such as dependencies or incompatibilities. That is, unlike traditional models of information systems (considering a single product), models of SPLs capture the variability among a family of products. 

\emph{Feature Models} (FMs) \cite{Benavides2010} are a popular family of notations, capable of describing complex SPLs. FMs can be constructed for a given application domain, using a methodology known as \emph{Feature-Oriented Domain Analysis} (FODA) \cite{Kang1990}. The output of this process is a FM, a diagram describing a complete SPL as a set of features and relationships. The diagram looks like a connected graph with the boxes/nodes in the diagram representing \textit{Features}, edges representing relationships, and cross-tree constraints expressed as Propositional logic formulas. A product becomes a \emph{configuration} of this diagram, i.e. a subset of features that satisfies all the relationships and constraints. Throughout this paper we will use \textit{product configuration}, \textit{product} and \textit{configuration} with interchangeable meanings.

\begin{figure}
   \caption{The vending machine FM created in FeatureIDE \cite{Kastner2009}.} 
   \includegraphics[width=\linewidth]{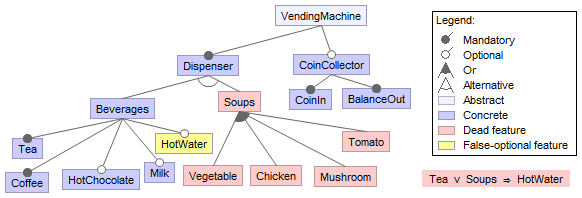}
   \label{fig:vending}  
\end{figure}

Given an FM, it is important to detect potential \emph{defects}, e.g. no valid product, \emph{dead features} (that cannot appear in any product), \emph{false optional features} (that are required in every product), etc. For instance, the FM illustrated in Fig.~\ref{fig:vending} is the model of a vending machine product line. A cross--tree constraint makes \textit{HotWater} a required feature when the features \textit{Tea} or \textit{Soups} are selected in the FM. This makes \textit{Soups} unselectable in any configuration as the \textit{Dispenser} can either have \textit{Beverages} or \textit{Soups} but never both. Meanwhile, optional feature \textit{HotWater} is required by mandatory feature \textit{Tea} (making it mandatory).

There are several techniques and tools that have been considered for this purpose, as surveyed in \cite{Benavides2010}. Among these tools, a popular choice is Alloy \cite{Jackson2002}, a light-weight formal modeling language with an efficient model finder. The generic semantics and versatility of the relational logic used in Alloy together with a fully automated analysis \cite{Torlak2006} makes Alloy an ideal choice for building complex automated tools based on it. Other options include \textit{Description Logic based} and \textit{Constraint programming based} approaches as illustrated in the survey by Benavides et al. \cite{Benavides2010}. 

The differences among the existing research works using Alloy lie in the formalization of the FM. Different formalizations provide unique benefits, e.g. readability or efficiency of the analysis. However, there is a lack of information regarding the relative merits of each proposal, making it difficult to select the most suitable one for a specific type of analysis. In this paper, we review the state of the art in the analysis of FMs using Alloy. Our goal is to characterize the strengths and weaknesses of various Alloy encoding approaches from several perspectives such as expressivity of the FM, efficiency of the analysis, existence of tool support, etc.

\par We present our findings over the following sections. Section~\ref{sec:Background} briefly discusses the related surveys and technical details about the Alloy analyzer and the FM analysis operations. Section~\ref{sec:ReviewMethod} presents the selection criteria used to locate the relevant works and the assessment criteria used to evaluate them. 
Then, Sections~\ref{sec:Evaluation} and ~\ref{sec:Discussion} discuss the results of the evaluation. Section~\ref{sec:Challenges} describes the major challenges identified in our survey with an indication of future research directions and Section~\ref{sec:Conclusion} summarizes the conclusions of this survey. 

\section{Background} \label{sec:Background} In this section we briefly discuss  the existing surveys on this research subject. Then, we present a short summary of the required technical know--how about Alloy and FM in general with an introduction to the benchmark FM. The section concludes with a short overview of the specific FM analysis operations that are considered by various research papers.

\subsection{Related surveys} There are 3 surveys \cite{Benavides2010,Lesta2015,Thum2014} which have systematically studied the state-of-the-art on FM analysis. Most of the high level findings related to tools support and capability of executing analysis operations have been documented in those surveys, but with no specific focus towards using Alloy as the analysis tool. Benavides et al. \cite{Benavides2010} have consolidated a complete list of analysis operations which are fundamentally required for a fully operational end-to-end FM analysis automated tool. We have used this list as our reference to evaluate the support levels extended by the different papers that are reviewed during this survey.

\subsection{Benchmark feature model} 
\begin{figure}[!b]    
   \includegraphics[width=\linewidth,height=12cm,keepaspectratio]{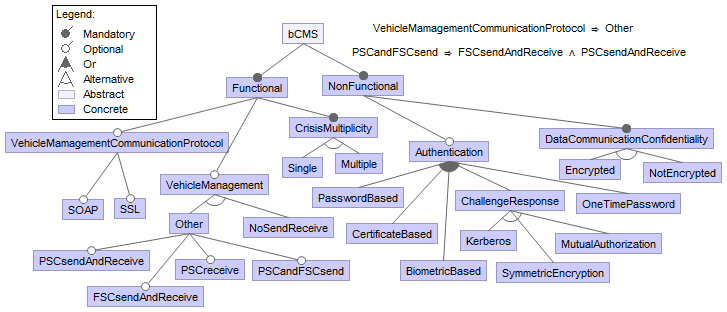}
   \caption{The bCMS FM \cite{Salay2014} created using FeatureIDE \cite{Kastner2009}.}
   \label{fig:bCMS}   
\end{figure}

For comparing the selected research works, a suitable FM has to be set as the benchmark. We have used the bCMS software product line \cite{Capozucca2012,Salay2014} (see Fig.~\ref{fig:bCMS}) for this purpose. The bCMS-SPL is a case study defined for the Workshop on Comparing Modeling Approaches (CMA'2011). It describes a family of car crash emergency systems in a FM with 28 features and 30 relationships among features. Two of these relationships are simple (at most 2 features involved) cross--tree constraints. 

\subsection{Alloy semantics} In Alloy, a model is described as a collection of \emph{signatures} \textbf{(\texttt{\textcolor{blue}{sig}})}, which identify the potential types for objects. Signatures can have \emph{fields} with values like references to other objects, sets of objects or mappings among objects. \emph{Facts} \textbf{(\texttt{\textcolor{blue}{fact}})} are constraints that capture the well-formedness rules of the model. Then, it is possible to check two types of properties: \emph{assertions} \textbf{(\texttt{\textcolor{blue}{assert}})}, i.e. searching an instance that violates a condition, or \emph{predicates} \textbf{(\texttt{\textcolor{blue}{pred}})}, i.e. searching an instance that fulfills it. When defining complex facts, predicates or assertions, it is possible to use \emph{functions} \textbf{(\texttt{\textcolor{blue}{fun}})} to reuse and encapsulate large subexpressions. 

Properties can be expressed using a relational logic that combines features from first-order logic (quantifiers and Boolean operators), set operations (navigations through references and mappings) and the operators of relational calculus (e.g. computing the differences among two relations).

\subsection{Analysis operations} In order to evaluate each encoding and the analysis operations supported by them, we have listed all the operations as described by Benavides et al. \cite{Benavides2010} across 2 tables: Table~\ref{table:FMAnalysis} (which lists out only those operations which are included in one or more research works considered in the survey) and Table~\ref{table:FMAnalysisCont} in Appendix~\ref{App:Appendix} (which enlists all the remaining analysis operations which are significant as mentioned by Benavides et al. \cite{Benavides2010}, but not covered by any research work, both among the ones considered in our review and in the larger space of research works related to FM analysis using Alloy as the model finder). More explanations on the importance and relevance of each analysis operation can be found in \cite{Benavides2010}.

\begin{table}[h!]
\begin{center}
\small
\captionof{table}{FM analysis operations commonly found across multiple research works.}
\label{table:FMAnalysis}
\noindent\resizebox{\textwidth}{!}{
\begin{tabular}{|p{0.5cm}|p{2cm}|p{11.5cm}|p{2.5cm}|}
\hline
\textbf{Id} & \textbf{Operation} & \textbf{Analysis operation description} & \textbf{Supported by} \\ \hline
A1	& Void FM & A feature model is void if it represents no products & \cite{Wang2005,Engineering2010,Huang2013,Ripon2012}	\\ \hline
A2	& Valid product	& A product is valid if it belongs to the set of products defined by the FM	& \cite{Ajoudanian2014,Jaime2012,Finkel2011,gheyi06,Huang2013,Engineering2010,Ripon2012,Tanizaki2008,Wang2005} \\ \hline
A3    & All products	& All valid products possible from the FM & \cite{gheyi06}	\\ \hline
A4    & Filter	& For a given partial product configuration this analysis will result in a list of valid products that can be generated from this partial configuration & \cite{Ajoudanian2014}	\\ \hline
A5    & Dead features	& A feature is dead if it cannot appear in any of the products of the SPL & \cite{gheyi06,Finkel2011}	\\ \hline
A6    & Wrong cardinalities	& These appear in cardinality-based feature models where cross-tree constraints are involved. A group cardinality is wrong if it cannot be instantiated	& \cite{Ajoudanian2014} \\ \hline
A7    & Refactoring & An FM is a refactoring of another one if both represent the same set of products while having a different structure	& \cite{Wang2005} \\ \hline
A8    & Commonality	& This operation takes an FM and a configuration as input and returns the percentage of products represented by the model including the input configuration & \cite{Ajoudanian2014} \\ \hline
A9    & Variability factor	& This operation takes an FM as input and returns the ratio between the number of products and 2\textsuperscript{n} where \texttt{n} is the number of features considered.	& \cite{Kang1990} \\ \hline
A10    & Degree of orthogonality & This operation takes an FM and a subtree (represented by its root feature) as input and returns the ratio of the total number of products of the FM and the number of products of the subtree	& \cite{Ajoudanian2014} \\ \hline
\end{tabular}
}
\end{center}
\end{table}

\section{Review method} 
\label{sec:ReviewMethod}

The review begins by considering existing surveys (e.g. \cite{Benavides2010,Thum2014,Lesta2015}) and searching the following terms using a logical \textit{and} operator for the combination, in the abstract and keyword lists of the following scientific databases:
\begin{center}
\begin{tabular}{l@{~~}l}
\textbf{Search terms:}  & Feature Model, Alloy, analysis, software product line, verification \\
\textbf{Data sources:} & IEEE Xplore, ACM Digital Library, ISI Web of Knowledge \\
\end{tabular}
\end{center}
The search process is iterated on the list of bibliographic references in each of the identified papers, in order to identify any potentially missing references. Overall, this initial search produced 47 research papers. Fig.~\ref{fig:plot} presents a timeline of these references while Table~\ref{table:ResearchPapers} in Appendix~\ref{App:Appendix} identifies their venue of publication. 

\begin{figure}
	\includegraphics[width=\linewidth,height=6cm,keepaspectratio]{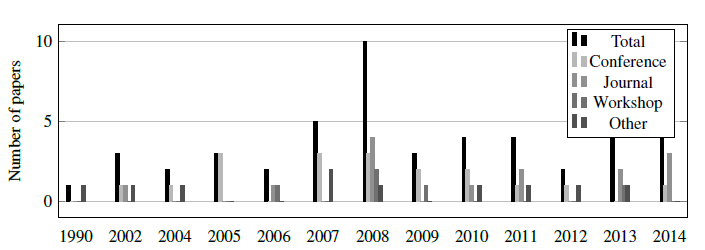}
   \caption{Collected papers statistics.}
   \label{fig:plot}
\end{figure}

In order to refine this collection of references, Table~\ref{table:Filtering} describes the inclusion and exclusion criteria used to select only the specific works related to our survey interest. On one hand, we focus on works dealing with the analysis of FMs following the \emph{Feature-Oriented Domain Analysis} (FODA) conventions \cite{Kang1990}. 
On the other hand, methods that include information beyond Feature Models, e.g.  behavioral models \cite{Dietrich2012} or use-case models \cite{Alfrez2014}, are omitted. As it is shown in Table~\ref{table:Filtering}, we only consider papers using Alloy as the verification engine. Thus, we omit any paper that uses alternative formalisms (e.g. constraint programming or description logic) or any other propositional logic provers \cite{Benavides2010,Lesta2015}. 

\begin{table}
\centering
\caption{Selection criteria used in this literature review.}
\label{table:Filtering}
\noindent\resizebox{\textwidth}{!}{
\begin{tabular}{|p{8cm}|p{8cm}|}
\hline
\textbf{Inclusion criteria} & \textbf{Exclusion criteria}\\
\hline
1. Discusses FM properties that can be analyzed using Alloy & 1. Does not consider FODA-style FMs describing a SPL \\
2. Focuses on Alloy-based Feature Model analysis of SPLs & 2. Does not consider the analysis of the FM \\
3. Introduces a new FM encoding or improve/extend existing encoding to include a new analysis & 3. Does not use Alloy as the underlying verification engine \\ \hline
\end{tabular}}
\end{table}
\normalsize

After applying the inclusion and exclusion criteria from Table~\ref{table:Filtering}, the final set of selected papers consists of 9 research papers which qualified to be included for our detailed review process. These papers from 2005 until 2015 include: 6 papers from conference proceedings \cite{Wang2005,Tanizaki2008,Engineering2010,Huang2013,Ripon2012,Jaime2012}, 2 from journal publications \cite{Finkel2011,Ajoudanian2014} and 1 technical report \cite{gheyi06}. 

\subsection{Threats to validity} \label{subsec:threats}
Before we go on further with the details of this literature survey, we would like to identify a set of potential weaknesses in our review process: 
\begin{itemize}
\item We have not included all possible list of literature databases, e.g. DBLP or Google scholar. On the other hand, the considered surveys \cite{Benavides2010,Thum2014,Lesta2015} are very recent (2010, 2014, 2015), and so they offer a good coverage of recent works. 
\item Other combinations of equivalent terms, such as \textit{"variability modeling"}, \textit{"automated FM analysis"}, etc. have not been considered. As a result of our search, and to the best of our knowledge, these nine selected papers are the most relevant for the topic addressed in this review. 
\item Marcilio et al. \cite{Marcilio2009} and Liang et al. \cite{Liang2015} have been successful in evaluating SAT-based FM analysis on very large FMs such as the FM of Linux kernel (5814 features), and found that the results were promising in terms of scalability of the FM and solver execution times. All such evaluations were done on FMs represented in Conjunctive Normal Form (CNF) which can be fed as input to SAT solvers such as the SAT4j standalone SAT solver. In this paper we could not use such very large FMs in CNF format mainly because we lacked the tool support for converting such CNF notations into Alloy specifications. Moreover it would be difficult to validate the results of analysis of various FM analysis operations on such large FMs. Therefore the scalability of Alloy encoding for larger FM specifications and its performance with respect to analysis execution time for such large FMs has not been covered in this paper.
\item We did not quantify the encoding performance in terms of the time taken to manually convert any FM to Alloy specification (ease of use).
\item As there is only one case study the respective merits of each encoding may vary on different FMs.
\end{itemize}

\section{Evaluation} \label{sec:Evaluation}

Our review was intended towards answering the following research questions for each selected work:

\begin{enumerate}
\item What is the goal of FM analysis, i.e. is there a motivating scenario?

\item What degree of expressivity is supported in the FM, i.e. what kind of relationships are allowed?

\item What kind of analysis can be performed on the FM? What is the efficiency of those analysis? 

\item What kind of tool support is available for generating the Alloy specification automatically?
\end{enumerate}

\subsection{Formalization strategies} This subsection provides an overview of the general formalization of FMs in the Alloy notation, in order to facilitate the comparison among the different formalization strategies. All selected works use common conventions in their Alloy encodings. For instance, all works share common concepts among the signatures they declare: \texttt{Feature}, \texttt{Relationship} and \texttt{Configuration}. A \texttt{Feature} represents a major functionality of a software product. \texttt{Relationships} are connections between features. Most connections are between a complex feature (\emph{parent}) and its set of sub-features (\emph{children}), defining a tree-like structure. However, it is possible to define \emph{cross-tree} relationships connecting arbitrary features, for example, stating that a feature  requires another feature. Finally, a \texttt{Configuration} (also called \texttt{Concept} in \cite{Wang2005}) is a set of features that define a product of an SPL.

\subsection{Goal of analysis} 
\begin{figure}[h]   
   \includegraphics[width=\linewidth]{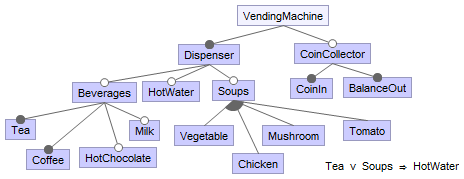}
   \caption{The corrected vending machine FM created in FeatureIDE \cite{Kastner2009}.} 
   \label{fig:corvending}  
\end{figure}

The goal of FM analysis in general should be to identify all sorts of problems (like dead features, presence of false optional features etc.) in the FM and suggest corrective measures to make the FM valid. For example, if we consider the \textit{Vending machine} FM shown in Fig.~\ref{fig:vending} it has both \textit{dead features} and \textit{false optional features}. Both these problems can be resolved if a \textit{'Refactoring' (A7)} FM analysis operation is applied and the structure of the FM is changed to Fig.~\ref{fig:corvending}. This FM is found to be completely valid with 162 possible product configurations (as calculated by FeatureIDE \cite{Kastner2009}). In our survey, every selected research work was targeting to emphasize on some or the other analysis operation on the feature model. The most common analysis operations are: check if an FM is void (A1) \cite{Wang2005,Engineering2010,Huang2013,Ripon2012}, finding invalid/valid configurations (A2) \cite{Ajoudanian2014,Finkel2011, Huang2013,Engineering2010,Wang2005,Ripon2012,Jaime2012,gheyi06,Tanizaki2008} and finding dead features (A5) \cite{Finkel2011,gheyi06}. The support extended for each analysis operation by the different selected papers in our review, has been marked under the \textbf{Supported by} column of Table~\ref{table:FMAnalysis}.

\subsection{Degree of expressiveness} 
Expressivity has been evaluated in terms of the ability to encode all possible relationships and constraints between features in the Alloy notation. We have consolidated all such relationships and constraints and tabulated them under the \textbf{Relationships} sub--table of Table~\ref{table:ReviewSummary}. This table shows that some papers lack support for some types of relationships in the FM. This was purely a choice of the respective authors, considering only the properties of interest and a trade-off between expressivity and efficiency. 

\subsection{Efficiency of the analysis}

To measure the efficiency, we have compared the execution time for various analysis operations using version 4.2 of the Alloy Analyzer (using the default SAT4j solver \cite{SAT4J2010}). We have used the Alloy encoding proposed by each paper to model the bCMS FM and then performed different analysis operations. Also note that we have not included any evaluation of the analysis performance compared to other verification engines beyond the Alloy Analyzer.

The details of the execution time in milliseconds can be found in Table~\ref{table:ReviewSummary}. 

\subsection{Available tool support}

Regarding the source of the FMs, there is no standard textual format to represent FODA style feature models. Hence, all works start from a manually created FM. In some cases \cite{gheyi06,Tanizaki2008,Huang2013,Ajoudanian2014,Jaime2012,Finkel2011}, the FM is a formal specification created specifically for analysis purposes. In others \cite{Wang2005,Engineering2010,Ripon2012}, the FM is the manual translation of the requirements document of a SPL. 

Regarding the generation of the Alloy specification from the FM, the inference was again that there was a lack of tool support: only Nakajima et al. \cite{Engineering2010} have a prototype tool called FD-Checker that can generate the Alloy specification from a propositional logic formula. All the other papers do not have any kind of tool support, so the Alloy formalization needs to be generated manually. In our opinion, this is a significant finding in this literature survey. This has also made it very difficult for us to evaluate the research results of the bCMS case study, as it involved manual translation of FM specifications to the respective Alloy encoding.

For each reference (1\textsuperscript{st} row) in Table~\ref{table:ReviewSummary} the set of supported relationships in the FM, i.e. their expressivity  (2\textsuperscript{nd} row), the availability of tool support for automatically generating the formal Alloy specification  (3\textsuperscript{rd} row) and the set of analysis operations offered by each method and its efficiency (4\textsuperscript{th} row) in terms of the execution time for each analysis, has been summarized.

\section{Discussion} \label{sec:Discussion}

In the following subsections we discuss the details of each encoding approach followed by the different research papers considered in this review in chronological order. Each research paper corresponds to one approach, also referred to as one strategy.

\subsection{Alloy encoding and analysis of bCMS FM}
In this subsection we include all major findings related to each selected paper and would provide the information based on the following criteria:
\begin{itemize}
\item Overview of the encoding strategy used - Specialty of the encoding and identifying the distinguishing aspect of the strategy.
\item Strengths and limitations of the encoding - Comparison with the encoding described by other research papers.
\item Ease of reproducing the research results - Discusses the effort to encode the bCMS FM and perform the analysis operations.
\item Major technical roadblocks - Discusses the technical limitations and roadblocks that were encountered while reproducing the analysis results on the encoded bCMS FM.
\end{itemize} 

\textbf{Wang et al. \cite{Wang2005}}: The encoding is thoroughly described in the paper. This is the only encoding which is able to perform a check for semantic equivalence between an FM and its refactored model. Unlike other papers this approach did not discuss about finding dead features, finding all possible product configurations or provide explanations for analysis results. The encoding was easy to reuse for the bCMS FM and the results stated in the paper were completely reproduced without any technical issues.

\textbf{Gheyi et al. \cite{gheyi06}}: Gheyi proposed two theories for Feature Model analysis, the R{-}Theory and the G{-}Theory. The R{-}Theory defined a FM as a set of features. 
The G{-}Theory was a concise approach towards creating reusable constructs that are generically implemented in order to allow easy FM specification in Alloy. The encoding was more mature than the specifications of Wang et al. \cite{Wang2005} in terms of expressivity as it was the only encoding among the surveyed papers that supported finding all valid products from any FM. It was easy to generate Alloy specifications for bCMS FM. FM analysis such as checking for valid/invalid product configurations, identifying dead features and collecting all valid product configurations was performed on bCMS FM and the results are presented in Table~\ref{table:ReviewSummary}. A major roadblock in the implementation of these theories in Alloy for the bCMS FM was the use of logical constructs (recursion) that are unsupported by the current version of the Alloy notation. This was resolved when we directly contacted the authors and got a solution from them.

\textbf{Tanizaki et al. \cite{Tanizaki2008}}: The encoding described in this paper is the most difficult to understand as the semantics used is very different from what we have seen in all the other approaches. Here the concept of a Feature-Model-Connection has been introduced in order to support traceability of configuration changes in a software system. The time taken to specify the bCMS FM was thrice the time what we took for the other encodings. The main roadblock was in understanding the encoding in terms of FM analysis as they were mainly targeted towards backtracking FM changes to identify any new invalid FM state. This is more or less a combination of analysis properties \textit{A12 \& A13}. But we were unable to reproduce the research results. This was mainly because of a lack of clarity in the encoding (too many signatures with no reusable functions or predicates which can aid in specifying bCMS FM in Alloy) encoding and incomplete encoding information in the paper (finding a product instance using the address book example from the paper was not directly reproduced because of the missing encodings which the authors have explicitly mentioned of having excluded it from the paper).

\textbf{Nakajima et al. \cite{Engineering2010}}: The most remarkable aspect of this encoding is that the features and its relationships are all encoded in terms of propositional logic formulas (using many \textcolor{blue}{\texttt{sig}} and \textcolor{blue}{\texttt{fact}}), rather than using the \textcolor{blue}{\texttt{fun}} or \textcolor{blue}{\texttt{pred}} constructs of Alloy. This makes the encoding effort for very large FMs a time consuming and error-prone process. The major strength of this encoding lies in its direct and simple to understand encoding semantics. But unfortunately this approach has a very long running time for the \textit{A2} analysis operation  (see Table~\ref{table:ReviewSummary}) even for such a small sized FM. Unlike Gheyi et al. \cite{gheyi06}, this encoding is not very compact while it is more flexible and potentially easier to be generated \textbf{automatically} for any given FM. For this reason they had introduced a tool called FD-Checker which could have been useful to convert propositional logic formulas to Alloy specifications of the FM. Nevertheless, the tool is currently not available in the public domain and we did not contact the authors as we were working on a smaller FM which we managed to encode manually.

\textbf{Finkel et al. \cite{Finkel2011}}: The goal of this paper was to detect all possible configurations and find if the model has any optional-feature flaw and missing-feature flaw. If a feature is marked as optional but exists in all valid products from the FM, then the FM is said to have an optional-feature flaw. Whereas, in the same way if a mandatory feature is absent in all products then the FM has a missing-feature flaw. For specifying the bCMS FM we had to manually encode the features as different signatures and specify relationships as propositional logic formulas in Alloy, similarly to Nakajima et al. \cite{Engineering2010}. Therefore the encoding was simple and easy to reproduce but it was time consuming and error-prone. This calls for automated tools to support the translation of FM to Alloy specifications.

\textbf{Ripon et al. \cite{Ripon2012}}: The main objective of this paper was to present an approach for formalizing and verifying SPL FMs with support for automatically generating customized products based on user requirements. They have used the same Alloy encoding as described by Gheyi et al. \cite{gheyi06} in their G-Theory and Wang et al. \cite{Wang2005}, and hence have similar findings. This can be seen from the results of analysis in Table~\ref{table:ReviewSummary}.

\textbf{Huang et at. \cite{Huang2013}}: The encoding is very similar to Nakajima et al. \cite{Engineering2010} as they have extended the same encoding to include signatures (\textcolor{blue}{\texttt{sig}}) that will enable analysis operations to check for valid/invalid sub--FMs in an FM. This approach is very verbose and error-prone due to relying completely on propositional logic formulas. Two analysis operations \textit{A1 \& A2} are successfully performed using this encoding on bCMS. Though the authors have even mentioned about detecting valid/invalid sub--models that aid in refining the overall FM using Refactoring (\textit{A7}), this was not explicitly demonstrated in the paper and hence it is not included in the bCMS analysis.

\textbf{Jaime et al. \cite{Jaime2012}}: The objective of this paper was to detect conflicts in the product configuration process which occur when intended features cannot be selected because of other choices of features included into the product. For this they have identified and included one more type of relationship which is \textit{'Non-selectable'} in addition to the relationship encodings provided by Gheyi \cite{gheyi06}. An FM is considered as invalid if full mandatory features that must be included in all the configurations are also non-selectable. The encoding was easy to understand and apply for specifying the bCMS FM. The analysis operation \textit{A2} took slightly more time compared to other approaches though the encoding was better encapsulated with appropriate predicates and functions for easy FM specification. There were no technical roadblocks.

\textbf{Ajoudanian et al. \cite{Ajoudanian2014}}: Constraint-based FMs with cardinalities are known as extended feature models. Hence they are feature models with attributes. This paper describes a promotion technique in Alloy which claims to significantly improve the efficiency of analysis operations performed on such extended Feature models. We were unable to reproduce the results of the paper using the example and Alloy specifications provided in it. When the Alloy encoding details were directly used, it showed several syntax errors in the Alloy editor such as missing \textbf{`\{'}, \textbf{`.'} between \textbf{\texttt{\textcolor{blue}{univ}}} and \textbf{`('}, undefined variables, invalid Alloy symbols for \textit{less than or equal} and \textit{greater than or equal}. Therefore it was difficult to reuse the encoding and reproduce their results with bCMS FM.

\subsection{Review summary} The evaluation summary is illustrated in Table~\ref{table:ReviewSummary}. After analyzing and comparing the execution time for different operations, we have identified the list of various encodings that can be used for the respective analysis operations based on the encoding efficiency and execution time. This is summarized in Table~\ref{table:Summary}.

				\begin{table}[t]
                \centering
                \captionof{table}{Review summary on the analysis of FM using Alloy.}
				\label{table:ReviewSummary}
\noindent\resizebox{\textwidth}{!}{
\begin{tabular}{|l|c|c|c|c|c|c|c|c|c|}
\hline 

\textbf{Reference}                                                                                               & \multicolumn{1}{c|}{Wang}                                            & \multicolumn{1}{c|}{Gheyi}                                              & \multicolumn{1}{c|}{Tanizaki}                                    & \multicolumn{1}{c|}{Nakajima}                                               & \multicolumn{1}{c|}{Finkel}                                   & \multicolumn{1}{c|}{Ripon}                                  & \multicolumn{1}{c|}{Huang}                                  & \multicolumn{1}{c|}{Jaime}                     & \multicolumn{1}{c|}{Ajoudanian}                     \\ 
& \multicolumn{1}{c|}{\cite{Wang2005}}                                            & \multicolumn{1}{c|}{\cite{gheyi06}}                                              & \multicolumn{1}{c|}{\cite{Tanizaki2008}}                                    & \multicolumn{1}{c|}{\cite{Engineering2010}}                                               & \multicolumn{1}{c|}{\cite{Finkel2011}}                                   & \multicolumn{1}{c|}{\cite{Ripon2012}}                                  & \multicolumn{1}{c|}{\cite{Huang2013}}                                  & \multicolumn{1}{c|}{\cite{Jaime2012}}                     & \multicolumn{1}{c|}{\cite{Ajoudanian2014}}                     \\ \hline

\textbf{Expressivity}                                                                                                  & R1-7 & R1-3, R5 & R1-3, R5-7 & R1-7 & R1-3, R5                             & R1-5                                                       & R1-3                                                       & R1-3, R5-7                                                       & R1-3, R5 \\ \hline
\textbf{Tool support}                                                                                                & None                                                   & None                                     & None                                                & FD-Checker                                                      & None                          & None                        & None                        & None                        & None \\ \hline
\begin{tabular}[c]{@{}l@{}}\textbf{Analysis}* \\ (Identifier + \\ Execution time)\end{tabular} & \begin{tabular}[c]{@{}c@{~~}r@{}}A1 & 5 ms\\ A2 & 7 ms\\ A7 & 1 ms\end{tabular}                 & \begin{tabular}[c]{@{}c@{~~}r@{}}A5 & 25 ms\\ A2 & 20 ms\\ A3 & 30 ms\end{tabular} & \begin{tabular}[c]{@{}c@{~~}r@{}} A2 & 10 ms\end{tabular}                      & \begin{tabular}[c]{@{}c@{~~}r@{}}A1 & 8 ms\\ A2 & 610 ms\end{tabular}                 & \begin{tabular}[c]{@{}c@{~~}r@{}}A2 & 7ms \\ A5 & 4 ms\end{tabular} & \begin{tabular}[c]{@{}c@{~~}r@{}}A1 & 10 ms\\ A2 & 8 ms\end{tabular} & \begin{tabular}[c]{@{}c@{~~}r@{}}A1 & 5 ms\\ A2 & 5 ms\end{tabular} & \begin{tabular}[c]{@{}c@{~~}r@{}} A2 & 22 ms\end{tabular} & \begin{tabular}[c]{@{}c@{~~}r@{}}A2 & 5 ms \end{tabular} \\  \hline 
\multicolumn{8}{l}{~*~Execution time of each analysis reported on the bCMS case study of Fig.~\ref{fig:bCMS}} \\
\multicolumn{8}{l}{~~~~Settings: PC with 16Gb RAM and a 3.4GHz processor. Alloy Analyzer version 4.2 (SAT4j SAT solver).} \\
\multicolumn{8}{l}{~~~~The Alloy source files for the bCMS product line, encoded using each approach are available at : } \\
\multicolumn{8}{l}{~~~~\href{https://www.dropbox.com/s/wysbls64ictoofm/Alloy.zip?dl=0}{https://www.dropbox.com/s/wysbls64ictoofm/Alloy.zip?dl=0}} \\
\end{tabular}}

\smallskip

\begin{tabularx}{\textwidth}{|@{~~}l@{~~}|@{~~}X@{~~}|}
\multicolumn{2}{l}{\textbf{Relationships}} \\
\hline
\textbf{Id}          & \textbf{Definition}                                                              \\ \hline
R1                  & \emph{Mandatory($x$):} \\&\quad
Feature $x$ must appear if parent($x$) is included \\ \hline
R2                   & \emph{Optional($x$) \textbf{or} Optional Or($x$):} \\&\quad Feature $x$ may or may not appear if parent($x$) is included \\ \hline
R3                & \emph{Alternative$(X)$ \textbf{or} XOR$(X)$}:\\&\quad Exactly one feature $x_i \in X$ must appear if parent$(X)$ is selected \\ \hline
R4       & \emph{Optional Alternative($X$):} \\&\quad
At most one feature $x_i \in X$ must appear if parent$(X)$ is selected
\\ \hline
R5                         & \emph{Or($X$):}\\&\quad At least one feature $x_i \in X$ must appear if parent$(X)$ is selected                                                                                \\ \hline
R6 & \emph{Requires($x,y$):} \\&\quad Feature $x$ must appear when feature $y$ is included                                                                           \\ \hline
R7  & \emph{Excludes($x,y$):} \\&\quad Feature $x$ and $y$ cannot appear both                                              \\ \hline
\end{tabularx}	

\smallskip

\end{table}

\begin{table}[!]
\centering
\caption{Analysis operation and best encoding strategy based on lowest execution time and highest expressiveness.}
\label{table:Summary}
\begin{tabular}{ll}
        \begin{tabular}{|p{3.5cm}|p{3cm}|}
        \hline
        \textbf{Analysis operation} & \textbf{Encoding strategy} \\ \hline
        A1 (Void FM) & Wang et al. \cite{Wang2005}  \\ \hline
        A2 (Valid product) & Wang et al. \cite{Wang2005}   \\ \hline
        A3 (All products) &  Gheyi et al. \cite{gheyi06}  \\ \hline
        A4 (Filter) & \textcolor{red}{Supported in \cite{Ajoudanian2014}*}   \\ \hline
        A5 (Dead features) & Gheyi et al. \cite{gheyi06}   \\ \hline
        \end{tabular}
        
        &

        \begin{tabular}{|p{3.5cm}|p{3cm}|}
        \hline
        \textbf{Analysis operation} & \textbf{Encoding strategy} \\ \hline
        A6 (Cardinalities) & \textcolor{red}{Supported in \cite{Ajoudanian2014}*}  \\ \hline
        A7 (Refactoring) &  Wang et al. \cite{Wang2005}  \\ \hline
        A8 (Commonality) &  \textcolor{red}{Supported in \cite{Ajoudanian2014}*}  \\ \hline
        A9 (Variability) & \textcolor{red}{Supported in \cite{Ajoudanian2014}*}   \\ \hline
        A10 (Orthogonality) &  \textcolor{red}{Supported in \cite{Ajoudanian2014}*}  \\ \hline
        \end{tabular}
\end{tabular}
\\       
\resizebox{\textwidth}{!}{\tiny{\textcolor{red}{*} These are discussed by Ajoudanian et al. \cite{Ajoudanian2014} but we were unable to reproduce the results due to various reasons as explained in Sec.~\ref{sec:Discussion}}}
\end{table}
\normalsize

We have identified two types of shortcomings in this literature survey that dealt with papers dealing with the analysis of FMs using Alloy: \emph{theoretical limitations} and \emph{practical limitations}.

\medskip \noindent
\textbf{Theoretical limitations.} Most encodings only support parent-child relationships and trivial requires or excludes cross-tree constraints. Furthermore, only one of the papers demonstrated the encoding to compute the entire set of products in the product line by extending support for checks such as counting the number of valid configurations. There is a wide scope for improving the encodings to allow more expressivity so that the other missing analysis operations (among the list of 30 as shown in Table~\ref{table:FMAnalysisCont}) can also be automated.

\medskip \noindent
\textbf{Practical limitations.} There were several challenges in our attempts to replicate each approach in the bCMS case study.

Firstly, only Nakajima et al. \cite{Engineering2010} describe a tool to generate the Alloy specification automatically. Even though, the tool is still available, it is not available for download and use. Therefore, almost at all times during the review process, the Alloy encodings had to be generated manually.

Furthermore, the encoding proposed by Ripon et al. \cite{Ripon2012} is not fully described. This means that some significant encoding is not provided in the paper and is neither available online. Attempts to fill the missing gaps yielded results which did not conform to the expected output obtained for the same analysis in the rest of encodings.

Finally, none of the works provided a large-scale FM for experimentation. Even though some approaches \cite{Ajoudanian2014} claimed being validated on FMs with hundreds of features, the examples are not available and authors did not respond to inquiries.

\section{Challenges and Future Research} 
\label{sec:Challenges}

The major challenge lies in having future research works targeting to address the missing analysis operations (Table~\ref{table:FMAnalysisCont}) using Alloy analyzer. The analysis operations such as \textit{Valid partial configuration}, \textit{Dependency analysis}, \textit{Redundancies} and \textit{Commonality analysis} have great significance when it comes to the industrial application of FM analysis during real product development. With the available research results we cannot implement a fully operational FM analysis automated tool based on Alloy without these missing operations that are crucial for generating feature models with lesser defects. Furthermore, it would be challenging to study in detail, all the formal proposals for semantics of feature models and how Alloy could be used to support each one of those semantics. As a starting point we would refer to the works of Schobbens et al. \cite{Schobbens2006} and Amador et al. \cite{Amador2015}. From the summary of review it can be inferred that some of the existing research results can be reused for specific analysis operations. Future research must focus to resolve all the limitations identified in our review.

\section{Conclusions} 
\label{sec:Conclusion} 

The main contribution of our work is to highlight the importance of future research in the direction of supporting automated FM analysis using Alloy, by identifying the limitations in the current state-of-the-art research contributions. In this paper we have selected, analyzed and compared the most relevant papers dealing with this subject. Several previous works have proposed the analysis of FODA-style FMs through Alloy. All formalizations but \cite{gheyi06} are unable to reason about more than one product configuration.
Lack of tool support for converting informal FM specifications to respective encodings is one of the major limitations identified during this survey. There are other practical issues that hinder the industrial application of these research results such as: out-of-date tools and Alloy formats, several syntactic errors in the Alloy specifications provided in the papers, lack of complete descriptions for the Alloy theories, lack of large-scale examples and a large number of analysis operations which are still not supported using Alloy. All this calls for future research to facilitate an efficient and automatic analysis of feature-based requirement specifications using Alloy. 

\medskip \noindent

\textbf{Acknowledgements.} The authors would like to thank Rohit Gheyi for responding to inquiries about his work on encoding FMs in Alloy and the reviewers for their useful comments.

\bibliographystyle{eptcs}
\bibliography{FMSPLE2016}
\newpage
\appendix

\section{Appendix} \label{App:Appendix}

Table~\ref{table:FMAnalysisCont} shows the remaining analysis operations that are important to be performed on FMs as indicated by Benavides et al. \cite{Benavides2010}. Table~\ref{table:ResearchPapers} shows the sources of the search results of all the papers that were found during the start of this survey on which we had further applied the inclusion and exclusion criteria to filter out the reviewed 9 papers.

\begin{table}[h!]
\begin{center}
\small
\captionof{table}{Analysis operations not discussed by any Alloy related research work till date.}
\label{table:FMAnalysisCont}
\noindent\resizebox{\textwidth}{!}{
\begin{tabular}{|p{0.5cm}|p{2cm}|p{14cm}|}
\hline
\textbf{Id} & \textbf{Operation} & \textbf{Analysis operation description} \\ \hline
A11    & Valid partial configuration	& An incomplete product configuration which in this partial state does not introduce any contradictions and can be still extended into a complete product		\\ \hline
A12    & Count & The number of valid products possible from the FM  \\ \hline
A13    &	Conditionally dead features & A feature is conditionally dead if it becomes dead under certain circumstances (e.g. when selecting another feature) 	\\ \hline
A14    & False optional features	& A feature is false optional if it is included in all the products of the product line despite not being modeled as mandatory 	\\ \hline
A15    & Redundancies	& An FM contains redundancies when some semantic information is modeled in multiple ways	 \\ \hline
A16    & Explanations	& This operation takes an FM and an analysis operation as input and returns information about the reasons of why or why not the corresponding response of the operation?  \\ \hline
A17    & Corrective explanations	& A corrective explanation provides suggestions to solve a problem, usually once this has been detected and explained 	\\ \hline
A18    & Generalization	& An FM, \textbf{F}, is a generalization of another one, \textbf{G}, if the set of products of \textbf{F} maintains and extends the set of products of \textbf{G}  \\ \hline
A19    & Specialization	& An FM, \textbf{F}, is a specialization of another one, \textbf{G}, if the set of products of \textbf{F} is a subset of the set of products of \textbf{G}	\\ \hline
A20    & Arbitrary edit	& There is no explicit relationship between two FMs  \\ \hline
A21    & Optimization	& This operation is chiefly useful when dealing with extended FMs where attributes are added to features. In this context, optimization operations may be used to select a set of features maximizing or minimizing the value of a given feature attribute  \\ \hline
A22    & Core features	& This operation takes an FM as input and returns the set of features that are part of all the products in the SPL  \\ \hline
A23    & Variants & Variant features are those that do not appear in all the products of the software product line	 \\ \hline
A24    & Atomic sets & An atomic set is a group of features (at least one) that can be treated as a unit when performing certain analyses. The intuitive idea behind atomic sets is that mandatory features and their parent features always appear together in products and therefore can be grouped without altering the result of certain operations  \\ \hline
A25    & Dependency analysis	& This operation takes an FM and a partial configuration as input and returns a new configuration with the features that should be selected and/or removed as a result of the propagation of constraints in the model \\ \hline
A26    & Multi--step configuration &	 A multi--step configuration problem is the process of producing a series of intermediate configurations, i.e. a configuration path, going from an FM configuration to another  \\ \hline
A27    & Homogeneity	& A more homogeneous FM would be one with few unique features in one product (i.e. a unique feature appears only in one product) while a less homogeneous one would be one with a lot of unique features.	\\ \hline
A28    & ECR	& Extra Constraint Representativeness - This operation takes an FM as input and returns the degree of representativeness of the cross-tree constraints in the tree	\\ \hline
A29    & LCA	& Lowest Common Ancestor - This operation takes a feature model and a set of features as input and returns a feature that is the lowest common ancestor of the input features \\ \hline
A30    & Root features	& This operation takes an FM and a set of features as input and returns a set of features that are the root features in the model	\\ \hline
\end{tabular}
}
\end{center}
\end{table}

\begin{table}
\captionof{table}{Source of the papers obtained in the initial search.}
\label{table:ResearchPapers}
\noindent\resizebox{\textwidth}{!}{
\begin{tabular}{|l|c|p{10.5cm}|}
\hline
\textbf{Source} & \textbf{No:}& \textbf{Details} \\ \hline
\textbf{Conferences} & \textbf{19} & \begin{tabular}{p{10.5cm}} Abstract State Machines, Alloy, B, TLA, VDM, and Z (ABZ) \\ Asia-Pacific Software Engineering Conference (APSEC) \\ Int. Conf. on Advanced Information Systems Engineering (CAiSE) \\  Int. Conf. on Coordination models and languages (COORDINATION) \\ European Conf. on Modelling Foundations and Applications (ECMFA) \\ 15th Int. Conf. on Conceptual Structures (ICCS) \\ Int. Conf. on Engineering of Complex Computer Systems (ICECCS) \\ Int. Conf. on Formal Engineering Methods (ICFEM) \\  ACM Symposium on Applied Computing (SAC) \\ Brazilian Symposium on Software Engineering (SBES) \\ Int. Conf. on Software Product Lines (SPLC) \end{tabular} \\ \hline
\textbf{Journals} & \textbf{14} & \begin{tabular}{p{10.5cm}}ACM Computing Surveys \\ ACM Transactions on Software Engineering and Methodology \\ Artificial Intelligence for Engineering Design, Analysis and Manufacturing \\ Communications of the ACM \\
IET Software 
\\Information and Software Technology
\\ Journal of Computational Science
\\ Journal of Information Systems
\\ Journal of Knowledge Engineering and Soft Data Paradigms \\ Journal of Logical and Algebraic Methods in Programming\\ Journal of Software \\ Journal of Systems and Software\\ Journal of Universal Computer Science \\ \end{tabular} \\ \hline
\textbf{Workshops} & \textbf{4} & \begin{tabular}{p{10.5cm}}Annual IEEE Software Engineering Workshop\\ Int. Workshop on Analysis of Software Product Lines\\ Int. Workshop on Variability Modelling of Software-Intensive Systems\\ First Alloy Workshop\end{tabular} \\ \hline
\textbf{Others} & \textbf{10} & \begin{tabular}{p{10.5cm}} PhD Thesis\\ Technical reports\end{tabular} \\ \hline
\end{tabular}}
\end{table}

\end{document}